# GRAVITATIONAL STABILITY FOR A VACUUM COSMIC SPACE CRYSTALLINE MODEL

J. A. Montemayor-Aldrete[1], J. R. Morones-Ibarra[2], A. Morales-Mori[3], A. Mendoza-Allende[1], A. Montemayor-Varela[4], M. del Castillo-Mussot[1] and G. J. Vázquez[1].

1. Instituto de Física, Universidad Nacional Autónoma de México, Apartado Postal 20-364, 01000 México, D. F.

2. Facultad de Ciencias Físico-Matemáticas, Posgrado, Universidad Autónoma de Nuevo León, Apartado Postal 101-F, 66450 San Nicolás de los Garza, Nuevo León, México

3. Centro de Ciencias Físicas, Universidad Nacional Autónoma de México, Apartado Postal 139-B, 62191, Cuernavaca, Morelos. México.

4. Centro de Mantenimiento, Diagnóstico y Operación Iberdrola, Polígono Industrial, El Serrallo, 12100, Castellón de la Plana, España.

**ABSTRACT**

Using Heisenberg's uncertainty principle it is shown that the gravitational stability condition for a crystalline vacuum cosmic space implies to obtain an equation formally equivalent to the relation first used by Gamow to predict the present temperature of the microwave background from the matter density. The compatibility condition between the quantum and the relativistic approaches has been obtained without infinities arising from the quantum analysis or singularities arising from the relativistic theory. The action which leads to our theory is the least action possible in a quantum scheme. The energy fluctuation involved in the gravitational stabilization of vacuum space is $10^{-40}$ times the energy of the crystalline structure of vacuum space inside the present Universe volume.

# 1. INTRODUCTION

The modern concept of a physical quantity in science follows that of Maxwell, where every expression of a quantity consists of two factors or components. One is the chosen standard quantity technically called the unit, and the other is the number of units required to make up the physical quantity [1]. From this conceptual scheme two different lines have been developed. One, the most important from a practical point of view, attends to the necessity to provide the basic units for measurements used in science, technology and everyday life [2]. The other line is devoted to the philosophical search for a deeper foundation of physical constant; see for instance works due to Maxwell [1], Planck [3] and Heisenberg [4].

Heisenberg's proposal of a natural system of units of measurement based on universal constants is a very interesting one, and is justified as follows [4]: "The universal constants determine the scale of nature, the characteristic quantities that cannot be reducing to other quantities. One needs at least three fundamental units for a complete set of units. A unit of length, one of time and one of mass is sufficient to form a complete set. The theory of relativity is connected with a universal constant in nature, the velocity of light, $c$. The quantum theory is connected with another universal constant of nature, Planck's quantum of action, $h$. There must exist a third universal constant in nature, this is obvious for purely dimensional reasons. An only a theory which contains this third unit (constant) can possibly determine the masses and other properties of the elementary particles. Judging from our present knowledge of these particles the most appropriate way to introducing the third universal constant would be the assumption of a universal length the value of which should be roughly $10^{-13}$cm, that is some what smaller than the radii of light nuclei. When from such three units one forms an expression which in its dimensions corresponds to a mass, its value has the order of magnitude of the masses of the elementary particles".

Following Heisenberg's proposal, it is possible to consider that some new physical result obtained in a previous paper [5] requires further analysis. In particular the possibility that vacuum cosmic space could have a crystalline structure, with a lattice parameter $r_N = R_{OU}/10^{40}$ ($r_N$ similar to neutron radius) where $R_{OU}$ is the present Universe radius deserves a careful exploration. From our point of view such physical analysis requires the study of compatibility conditions between the General Theory of Relativity and Quantum Theory as applied to the crystalline vacuum cosmic space model. This because a theory which physically describes the metric of cosmic space and its evolution and a quantum theory which allows to determine the masses of elementary particles (neutrons) existing in such metric are simultaneously required to analyze such problem.

The main purpose of this paper is to study the immediate implications about the gravitational stability of a model which considers that the vacuum cosmic space has a crystalline structure with a lattice parameter of the order of the neutron radius $r_N \sim 10^{-13} cm$.

## 2. FORMALISM

Our model for the vacuum cosmic space is an infinite crystalline structure characterized by a lattice parameter roughly the size of the neutron radius which is the distance between the physical entities that form the crystalline structure. Here vacuum means, by definition, the state of lowest or minimum energy per unit volume. The state $|0>$ is the state of crystalline structure without deformation. According to Einstein's gravitational theory, in this crystalline structure there is a gravitational attraction between such physical entities, and also there appears an interaction between them due to crystalline lattice deformation. These phenomena lead to a gravitational instability of such vacuum cosmic space.

In nature we have a lot of systems which, in principle, are unstable against some kind of force but thanks to quantum forces which arise from Heisenberg's uncertainty principle the system gets stability. Here we mention two examples: the atom and the nucleus.

As we know, in the hydrogen atom, which classically is an unstable system, we can obtain stability by using Heisenberg's uncertainty principle, which introduces something like a compensatory quantum force, which stabilise, the system.

For the case of the hydrogen atom, where the Hamiltonian is given by $H = \dfrac{p^2}{2m_e} - k\dfrac{e^2}{r}$

$$E = \langle H \rangle = \dfrac{1}{2m_e}\langle p^2 \rangle - ke^2 \left\langle \dfrac{1}{r} \right\rangle,$$

where symbols have their usual meaning. Assuming that $\Delta p \approx p$, $\Delta r \approx r$ and from the requirement that $\dfrac{dE}{dr} = 0$ we obtain that the radius $R_m$ that minimises the energy is the Bohr radius

$$R_m = a_0 = \dfrac{\eta^2}{ke^2 m_e} \qquad (1)$$

The main point here is that, in contrast to classical mechanics, the energy is bounded from below because of the uncertainty principle.

Similarly, for the case of nuclear forces, in the deuteron the stability can be explained by using again Heisenberg's uncertainty principle, or in nuclei with several nucleons, the stability can be achieved appealing to this principle as Yukawa did in 1935, explaining the nuclear force by the particles exchange, through the relation

$$m_\pi c^2 = \dfrac{\eta c}{r_N} \qquad (2)$$

Following the same arguments as above, we can extend these ideas to a system which interacts by gravitational forces. In the same way that Heisenberg's uncertainty principle is appealed for stabilizing a system like the hydrogen atom or a light nucleus, in this work this principle is used to prevent the collapse of a crystalline structure, which is our model for the vacuum cosmic space, due to the action of gravitational stresses.

In our crystalline model of vacuum cosmic space with lattice parameter of the order of the neutron radius $r_N$, the number of physical entities which exist for such crystal inside a volume $V_{OU} = \frac{4}{3}\pi R_{OU}^3$, $N_{CVS}$ is $10^{120}$, provided that $R_{OU}$ is the radius associated to the most usual value for the Universe age: 15000 millions years. Einstein's gravitational theory states that such physical arrangement is unstable under the action of long-range gravitational stresses. We can restore the equilibrium or stability of this system around average gravitational stresses with zero value by using Heisenberg's uncertainty principle. Each of the $N_{CVS}$ entities inside $V_{OU}$ behaves as a linear harmonic oscillator. In general we have $3N_{CVS}$ degrees of freedom in this system for vibration modes [6], but due to the radial symmetry of the gravitation interaction we have only $N_{CVS}$ degrees of freedom, which correspond to $3N_{CVS}$ linear one dimensional harmonic oscillators. Each of the $N_{CVS}$ physical entities of the crystalline vacuum cosmic space inside the volume $V_{OU}$, contributes with a stabilisation energy $\Delta\varepsilon_{OU}$ against gravitational forces, given by

$$\Delta\varepsilon_{OU}\Delta t_{OU} \geq h, \qquad (3)$$

where $\Delta t_{OU} \equiv \frac{R_{OU}}{c}$, is the time that gravitational waves require to traverse the Universe's radius $R_{OU}$. Thus, Eq. (3) can be written as

$$\Delta\varepsilon_{OU} \geq \frac{hc}{R_{OU}} \qquad (4)$$

Now by using the relation $\lambda\nu = c$ and defining $\nu_{OU} \equiv \frac{c}{\lambda_{OU}} = \frac{c}{R_{OU}}$, Eq. (4) can be written as

$$\varepsilon_{OU} \equiv \Delta\varepsilon_{OU} \geq h\nu_{OU} \qquad (5)$$

Eq. (5) describes the fundamental quantum of gravitational waves which, in principle are responsible for the gravitational stability of the vacuum cosmic crystalline structure.

It is important to note that in the three cases that we have considered, the stabilization of fundamental physical systems against instabilities arising from electromagnetic forces, nuclear forces and gravitational forces which has led to Eqs. (1), (2) and (4), the stability radii are inversely proportional to the rest energy (self-energy) of the "particle" which is orbiting. For the three cases the De-Broglie matter theory [7] states that each "particle" orbiting around a radius $r$, satisfies the De-Broglie relation $\lambda(r) = \dfrac{h}{p}$, where $p$ is the momentum of the orbiting physical entity, circling in a stationary wave.

For low frequencies, the relation between absolute temperature, $T$, and the photoenergy [8, 9] is given by:

$$E_p = h\nu = kT, \qquad (6)$$

where $k$ is the Boltzman constant. In addition, for weak gravitational fields, which correspond to the linear region of the Einstein's equations, there is a strong analogy between Maxwell's and Einstein's equations, so electromagnetic and gravitational waves have a similar behaviour. We assume then that Eq. (6) is also satisfied by gravitational waves, leading to $kT_{OU} \geq h\nu_{OU}$, or equivalently,

$$T_{OU}\lambda_{OU} \geq \dfrac{hc}{k}, \qquad (7)$$

where $T_{OU}$ is the Kelvin temperature associated with gravitational waves with wavelength of the order of the present Universe radius, $R_{OU} \approx \lambda_{OU}$, which corresponds to the temperature $T_{OU} \cong 10^{-27} K$. Let us consider the relations $\Delta t_{OU} \geq \dfrac{\eta}{\Delta\varepsilon_{OU}} = \dfrac{\eta}{h\nu_{OU}} \approx \dfrac{1}{\nu_{OU}}$, combining the last result with Eq. (7), we obtain

$$\Delta t_{OU} T_{OU} \geq \dfrac{h}{k} \qquad (8)$$

4Eq. (7) for gravitational waves which stabilize the cosmic vacuum crystalline space (CVCS) resembles the Wien's displacement law for electromagnetic black body radiation [10]

$$\lambda_{max} T = \frac{hc}{k}\left(\frac{1}{4.96511423}\right) \quad (9)$$

The $N_{CVS}$ gravitational waves quanta required to stabilize the crystalline structure of vacuum space in a volume $V_{OU} = \frac{4}{3}\pi R_{OU}^3$, lead to an adiabatic compression process due to the gravitational attraction effect between them.

According to Peebles [11] during an adiabatic expansion of gravitational waves, the fractional change in the frequency $\frac{\Delta \nu}{\nu}$ and the fractional change in the radius $\frac{\Delta r}{r}$ of the volume enclosing the gravitational waves, are related through

$$\frac{\Delta \nu}{\nu} = -\frac{\Delta r}{r} \quad (10)$$

The same expression applies to the adiabatic gravitational compression process derived by their own gravitational attraction.

For an isentropic process of expansion by electromagnetic radiation [12], we have

$$T^3 V = constant, \quad (11)$$

where $V$ is the cavity volume containing the electromagnetic radiation. For an spherical cavity of radius $r$, $Tr = constant \equiv c_o$. So,

$$dr = -\frac{c_0}{T^2} dT \quad (12)$$

From Eq. (10) and Eq. (12), $\frac{d\nu}{\nu} = \frac{dT}{T}$. Integrating this expression, gives,

$$T\lambda = constant \equiv c_2 \quad (13)$$

Applying for $T_{OU}$, becomes $T_{OU} \lambda_{OU} = c_2$, then by comparing with Eq. (7) a value for $c_2$, is obtained $c_2 \geq \dfrac{hc}{k}$. So, in general,

$$T\lambda \geq \frac{hc}{k} \qquad (14)$$

Or equivalently,

$$\frac{T}{\nu} \geq \frac{h}{k} \qquad (15)$$

Eq. (14) is a generalization of Eq. (7).

But Eq. (14) appears in a new physical-geometrical aspect by considering the De Broglie equations and its geometrical meaning; if we use the relation $n\lambda(r) = 2\pi r$ into Eq. (14) the gravitational waves temperature is then given by the following expression:

$$T(r) \geq \left(\frac{hc}{k}\right) \frac{n}{2\pi r} \qquad (16)$$

If this equation is applied for $r = R_{OU}$ and for $r = r$ it is clear that the following equation is obtained,

$$T(r)\, r = T_{OU}\, R_{OU} \qquad (17)$$

or

$$\frac{T(r)}{T_{OU}} = \frac{R_{OU}}{r} \qquad (18)$$

During the adiabatic gravitational wave compression process, the total energy $E_{OU}(R_{OU}) = 10^{120} \varepsilon_{OU}$ of the $N_{CVS}$ gravitational quanta required to stabilise the CVCS of volume $V_{OU} = \frac{4\pi}{3} R_{OU}^3$ remains constant. Due to Energy conservation the energy densities of these gravitational waves, $U$, are related through the expression $U_{OU} R_{OU}^3 = U_U(r) \, r^3$. Or, equivalently

$$\frac{R_{OU}}{r} = \left(\frac{U_U(r)}{U_{OU}}\right)^{\frac{1}{3}} \qquad (19)$$

which combined with Eq. (18) gives

$$\frac{T(r)}{T_{OU}} = \left(\frac{U_U(r)}{U_{OU}}\right)^{\frac{1}{3}} \qquad (20)$$

Eq. (20) resembles an equation previously obtained by Gamow. According to Penzias [13]: "once pair production has ceased $\rho$, the matter density, varies simply as

$$\frac{T_1}{T_0} = \frac{L_0}{L_1} = \left(\frac{\rho_1}{\rho_0}\right)^{\frac{1}{3}} \qquad (21)$$

(Where $T_1$ and $T_0$ are absolute temperatures, $L_1$ and $L_0$ are radial distances). If we take $T_1$ and $\rho_1$ to be the radiation temperature and matter density at the time of deuterium formation ($10^9 K$ and $10^5 \frac{g}{cm^3}$), we have the relation first used by Gamow to predict the present temperature of the microwave background from the density of matter".

The resemblance between Eqs. (20) and (21) is evident. However Eq. (20) refers to a physical situation of an adiabatic process of compression of gravitational waves under their own interaction, whereas Eq. (21) refers to the adiabatic expansion of matter after the Big Bang. In fact, the physical process which leads to Eq. (20) explains the possibility of the Big Bang event without singularities, as a transformation process of the fundamental gravitational quanta, mentioned before, into matter quanta.

## 3. DISCUSSION AND CONCLUSIONS

3.1. By using Heisenberg's uncertainty principle it has been shown that the vacuum cosmic space could be gravitationally stable. The model for such vacuum space is crystalline with a lattice parameter of the order of the neutron radius and the volume used to obtain gravitation stability from the collective quantum fluctuations is about the present Universe volume. Then Heisenberg's uncertainty principle allows to stabilise not only microscopical systems against electromagnetic or nuclear forces but also to stabilise macroscopically system against gravitational forces.

3.2. The big bang would have resulted from a previous adiabatic compression process of gravitational waves. In other words, from the analysis of the adiabatic process of compression between the gravitational waves, which stabilises the vacuum cosmic space with crystalline structure, an equation which describes the relationship between temperature, radial distance and gravitational wave energy density has been obtained, Eq. (20). Such an equation is formally equal to the equation used by Gamow's team to predict the present microwave temperature background from the density of matter, Eq. (21). This last equation has been obtained by Gamow from the General Relativity Theory applied to the big bang event, in particular arising from the Friedmann-Robertson-Walker Equation. But Eq. (20) is also formally identical to a previous result obtained by Homer Lane in 1869, called by Chandrasekhar "The Lane's Theorem" [14].

In a global way, by using energy conservation, we show that $10^{120}$ quantum of gravitation energy, each one with an energy $h\nu_{OU} = hc/R_{OU}$, becomes into $10^{80}$ neutrons as required by the Gamow analysis previously mentioned. By using the energy conservation principle it is possible to show that the diminishing in the gravitational energy of the crystalline gravitational field during the adiabatic compression of the gravitational waves from $R_{OU}$ to the radius which envelopes the $10^{80}$ neutrons formed at the end of the contraction phase is the energy source required to produce an electromagnetic radiation with a total energy of $10^{13} E_{OU} = 10^{93} u_N$. This physical consideration gives the conditions for a hot big bang in our theoretical scheme. The expansion cycle will be treated at detail in a future paper.

3.3. Conditions for compatibility between the quantum analysis of the gravitational stability of the vacuum space and the relativistic analysis of the big bang has been obtained without infinities arising from the quantum analysis or singularities appearing from the relativistic theory.

3.4. Our theoretical analysis is in agreement with the least action principle but also it is not possible, by theoretical construction, that any other model exhibits a least action than our model. The required action for our model is more o less equal to that involved in the Gamow approach.

3.5. At difference of the big bang theory, our scheme is a theory with initial conditions; this characteristic opens up the possibility that its predictive power will be greater than the big bang theory.

3.6. In our crystalline structure scheme of vacuum cosmic space a relativistic theory for the big bang does not violate the energy conservation principle; but the standard big bang theory does. This is because on the one hand, in our scheme the crystalline vacuum cosmic space is an eternal structure and the energetic fluctuation involved in the gravitational stabilisation of each volume $V_{ou}$ is about $10^{-40}$ of the energy per unit volume of the crystalline vacuum cosmic space which is compatible with Heisenberg's uncertainty principle. And on the other hand, in the standard theory of the big bang the vacuum cosmic space is growing together with the Universe expansion. The Friedman - Robertson - Walker equation obeys the energy conservation by neglecting the quantum energy arising from the increase of the vacuum cosmic space volume, which for cosmological volumes is a huge quantity.

133.7. It is clear that in our scheme the only energy which enters in the Einstein's field equations is due to the long - range quantum fluctuations of the crystalline vacuum cosmic space in the form of gravitational waves, electromagnetic energy and matter and antimatter.

3.8. This work links a quantum analysis about collective interactions between all the elements of a macroscopically system with a non - quantum relativistic cosmological model which has an objective physical reality. In our analysis an objective quantum picture of the crystalline vacuum cosmic space arising from the long - range interaction between their lattice entities appears. This conclusion contradicts the Neils Bohr and Stephen Hawking consideration (among few other) about that there is no objective picture at all, and which considers that: actually there is nothing "out there" at the quantum level. Some how, reality emerges only in relation to the results of "measurements" made by human beings. Quantum theory, according to this view, provides merely a calculation procedure and does not attempt to describe the world as it actually "is". Following Penrose's [15] and Barrow's analysis [16] it is clear that both Bohr's and Hawking's analysis mislead the point by confusing between "measurement" made by human observers and interaction. Interactions which occur between any physical entities all the time in all the places of the Universe governed by Heisenberg's uncertainty principle as has been shown here and in many other applications do no require the existence of human beings to take place. In other words, according to our model the physical reality at quantum level as applied to the so called Universe does not require the presence of a conscious being to exist.

3.9. There are many problems which remain to be solved in our theoretical scheme for instance: the thermodynamical aspects related to the formation of quantum matter packages, the evaluation of the cosmological constant implied by our model, the very low entropy value at the starting of the big bang, the quantum aspects of the formation process of the quantum matter packages, the relation between the matter and antimatter production, etc. All these problems will be addressed in further contributions.

## ACKNOWLEDGEMENTS

We want to specially thank Prof. M. López de Haro for many years of deep discussions and arguments, for his contribution to final shaping of the ideas and for his encouragement not to give up and unorthodox approach to cosmological problems and also we acknowledge to the librarian Technician G. Moreno for her stupendous work and to O. N. Rodríguez Peña for her patient and skilful typing work.